\documentclass[aps,prl,twocolumn,floatfix,10pt]{revtex4-1} 
\usepackage{amsmath,amssymb,amsfonts,amsthm,url,xspace,graphicx,psfrag}
\usepackage[pdftitle={XOR-Ising Loops and the Gaussian Free Field},
            pdfauthor={David B. Wilson},
            bookmarks=true,bookmarksopen=true,bookmarksopenlevel=2,
            pdffitwindow=true,
            pdfmenubar=false,
            pdftoolbar=false,
            hypertex]{hyperref}
\usepackage[all]{hypcap}
\usepackage{multirow}

\makeatletter
\def\url@leostyle{%
  \@ifundefined{selectfont}{\def\UrlFont{\sf}}{\def\UrlFont{\small\ttfamily}}}
\makeatother
\newcommand{\arXiv}[1]{\href{http://arxiv.org/abs/#1}{arXiv:#1}}
\newcommand{\arxiv}[1]{\href{http://arxiv.org/abs/#1}{#1}}
\newcommand{\sref}[1]{\hyperref[#1]{\S~\ref*{#1}}}
\newcommand{\aref}[1]{\hyperref[#1]{Appendix~\ref*{#1}}}
\newcommand{\lref}[1]{\hyperref[#1]{Lemma~\ref*{#1}}}
\newcommand{\tref}[1]{\hyperref[#1]{Theorem~\ref*{#1}}}
\newcommand{\cref}[1]{\hyperref[#1]{Corollary~\ref*{#1}}}
\newcommand{\fref}[1]{\hyperref[#1]{Figure~\ref*{#1}}}
\newcommand{\pref}[1]{\hyperref[#1]{Proposition~\ref*{#1}}}

\newcommand{\tbref}[1]{\hyperref[#1]{Table~\ref*{#1}}}

\newcommand{\MRhref}[2]{\href{http://www.ams.org/mathscinet-getitem?mr=#1}{MR#2}}

\makeatletter
\def\@strippedMR{}
\def\@scanforMR#1#2#3\endscan{%
  \ifx#1M\ifx#2R\def\@strippedMR{#3}%
  \else\def\@strippedMR{#1}}
\newcommand\MR[1]{\relax\ifhmode\unskip\spacefactor3000 \space\fi
  \@scanforMR#1\endscan
  \MRhref{\expandafter\@rst \@strippedMR other}{\@strippedMR}}
\newcommand\MRs[1]{\relax\ifhmode\unskip\spacefactor3000 \space\fi
  \@scanforMR#1\endscan
  \MRhref{\@strippedMR}{\@strippedMR}}
\makeatother

\numberwithin{theorem}{section}

\theoremstyle{definition}
\theoremstyle{definition}

\newcommand{\SLE}{\operatorname{SLE}}
\newcommand{\CLE}{\operatorname{CLE}}

\newcommand{\eps}{\varepsilon}

\newcommand{\old}[1]{}

\newcommand{\LL}{{\mathcal L}}

\def\rcs $#1: #2 ${\expandafter\def\csname rcs#1\endcsname {#2}}
\rcs $Date: 2011/02/16 03:14:17 $

\begin{document}
\title{XOR-Ising Loops and the Gaussian Free Field}
\author{\href{http://dbwilson.com}{David B. Wilson}}
\affiliation{\href{http://research.microsoft.com}{Microsoft Research}, Redmond, WA 98052, USA}
\date{February 18, 2011}

\begin{abstract}
  We find by simulation that the interfaces in the exclusive-or (XOR)
  of two independent 2D Ising spin configurations at the critical
  temperature form an ensemble of loops that have the same
  distribution as the contour lines of the Gaussian free field, but
  with the heights of the contours spaced $\sqrt{2}$ times as far
  apart as they are for the conformal loop ensemble $\CLE_4$ or the
  double dimer model on the square lattice.  For domains with
  boundary, various natural boundary conditions for the two Ising
  models correspond to certain boundary heights for the Gaussian free
  field.
\end{abstract}

\maketitle

\section{XOR-Ising loops and related models}

\begin{figure}[b!]
\psfrag{+}[cc][cc]{$\oplus$}
\psfrag{=}[cc][cc]{$=$}
\includegraphics[width=\columnwidth]{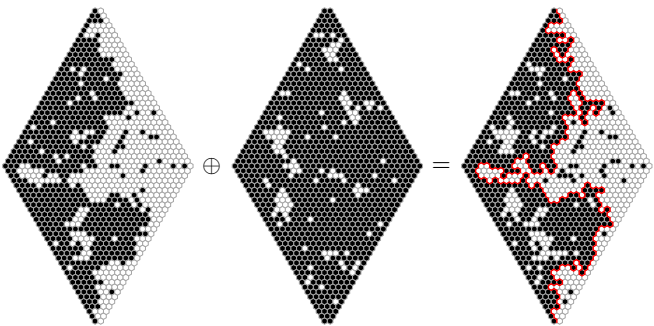}
\caption{(Color online)
  On the left is the critical Ising model on a lozenge-shaped
  domain of order $L=32$ with black-white boundary conditions.  The
  interface from bottom to top is in the scaling limit $\SLE_3$
  \cite{smirnov:ising,chelkak-smirnov}.  In the middle is the critical
  Ising model with black boundary conditions.  The ensemble of loops
  is in the limit $\CLE_3$ \cite{smirnov:ising,chelkak-smirnov}.  On
  the right is the XOR of the left and middle spin configurations.
  The 
  interface from bottom to top appears to be
  $\SLE_{4,\sqrt{2}-1,\sqrt{2}-1}$ in the scaling limit.}
\label{fig:xor-ising}
\end{figure}

The double-Ising model consists of two independent Ising spin
configurations $\sigma$ and $\tau$ on the same graph.  The
double-Ising model on planar graphs is closely related to a variety of
statistical physics models, including the Ashkin-Teller model
\cite{baxter,MR942673}, the 8-vertex model \cite{baxter,MR942673}, the
Gaussian model
\cite{kadanoff-brown,Knops1980448,hilhorst-leeuwen,dF-S-Z,boyanovsky,MR992372},
the double-dimer model, and spanning trees
\cite{boutillier-tiliere,tiliere}, some of which we discuss below.
The XOR-Ising configuration $\xi$ (illustrated in
\fref{fig:xor-ising}) is defined by $\xi_v = \sigma_v \times \tau_v$
for each vertex $v$ of the graph (if the spin values are $\pm 1$), or
equivalently, $\xi_v = \sigma_v \oplus \tau_v$ (if the spin values are
false and true), where $\oplus$ denotes exclusive-or (XOR).  The
XOR-Ising spins $\xi$ also called the polarization of the double-Ising
model \cite{kadanoff-brown}.  The interfaces between $+1$ spins and
$-1$ spins in an XOR-Ising configuration form loops or paths
connecting boundary points.  We find empirically that when the two
Ising models are at the critical temperature, the XOR-Ising loops are
closely related to contour lines of a Gaussian free field, but that,
in a sense that we will make more precise, there are fewer XOR-Ising
loops than loops in other loop models that are related to the Gaussian
free field.

\begin{table*}
\begin{ruledtabular}
\begin{tabular}{clll}
quantity & XOR-Ising loops & \ \ \ nearby exact value & $\CLE_4$ loops (exact value) \\
\hline
loop dimension & $1.5000 \pm 0.0003$ & \ \ \ \ $1.5000\dots=3/2$ & \ \ \ \ $1.5000\dots=3/2$ \\
winding angle variance $/ \log L$ & $1.0000 \pm 0.0003$ & \ \ \ \ $1.0000\dots=1$ & \ \ \ \ $1.0000\dots=1$ \\
E[\# loops around point]$/ \log L$ & $ 0.07168 \pm 0.00005 $ & \ \ \ \ $0.07164\dots=1/(\sqrt{2} \pi^2)$ & \ \ \ \ $0.10132\dots=1/\pi^2$ \\
Var[\# loops around point]$/ \log L$\ & $0.05057\pm 0.00003$ & \ \ \ \ $0.05066\dots=1/(2 \pi^2)$ & \ \ \ \ $0.06755\dots=2/(3\pi^2)$ \\
gasket dimension & $1.91400 \pm 0.00006$ & \ \ \ \ $1.91421\dots=1/2+\sqrt{2}$ & \ \ \ \ $1.87500\dots=15/8$ \\
electrical thickness & \ \ shown in Fig.~\ref{thickness} & excursion time of BM from $[-\pi,\pi]$\ \ \ & excursion time of BM from $[-\pi,\pi]$ \\
difference in log conformal radii & \ \ shown in Fig.~\ref{radii-diff} & exit time of BM from $[-\pi,\sqrt{2}\pi]$ & exit time of BM from $[-\pi,\pi]$ \\
\end{tabular}
\end{ruledtabular}
\caption{
  Comparison of XOR-Ising loops with the conformal loop ensemble $\CLE_4$.
  The (first five) measured values are from configurations on $L\times L$
  tori for $L=2^{10},2^{11},2^{12}$.
  The properties of individual XOR-Ising loops appear to be the same as
  for $\CLE_4$ loops, but the two loop ensembles are different.  The
  measured values of the XOR-Ising loops are consistent with contour
  lines of the Gaussian free field, but with height differences between
  contours being $\sqrt{2}$ times as large as for $\CLE_4$.
}
\label{tbl:params}
\end{table*}

XOR-Ising interfaces are (essentially) the same as the double-dimer
model on the Fisher lattice, as we now explain.  A dimer configuration
on a graph is a pairing of the vertices such that every vertex is
paired with exactly one of its neighbors.  Fisher showed that the
Ising model on any graph has the same partition function as dimer
configurations on a related graph \cite{fisher}.  This relation is not
combinatorial, but for \textit{planar\/} graphs there is a
combinatorial correspondence between dimers and Ising spins, based on
the Ising model's low temperature expansion \cite{baxter}.  When the
Ising spins are on the triangular lattice, the dimer configurations
are on a graph consisting of dodecagons and triangles, called the
Fisher lattice.

The double-dimer model is formed by superimposing two independent
random dimer configurations; the result is a collection of loops and
doubled edges.  If one of the dimer configurations has two defects
(monomers) on the boundary, then the double-dimer configuration also
has a path connecting the defects.  If we ignore the route that a
double-dimer loop (on the Fisher lattice) takes within the triangles,
the double-dimer loop traces out a path where one Ising configuration
has aligned spins while the other one does not.  Then double-dimer
loops on the Fisher lattice (ignoring doubled edges and the
intratriangle por-tions of loops) are just the XOR-Ising interfaces.

It is instructive to compare XOR-Ising loops with the double-dimer
model on other lattices.  On the square lattice or hexagonal lattice
with suitable boundary conditions, the scaling limit of a double-dimer
path connecting two boundary points is conformally invariant
\cite{kenyon:ddci} and is thought to be described by Schramm-Loewner
evolution with parameter~4 ($\SLE_4$) (see e.g.,
\cite{MR2153402,schramm:sle,MR2148644}).  The scaling limit of the
whole ensemble of loops is conjectured to be the conformal loop
ensemble with parameter~4 ($\CLE_4$) \cite{sheffield:cle}.  (The
conformal loop ensemble $\CLE_\kappa$ is thought to be the scaling
limit of the $O(n)$ model when $n=-2\cos(4\pi/\kappa)$
\cite{sheffield:cle}.)  Dimer configurations on the square and
hexagonal lattices have height functions that are known to behave like
the Gaussian free field in the scaling limit
\cite{MR1872739,MR2415464}, and $\CLE_4$ is known to be the scaling
limit of contour lines of the Gaussian free field where the heights of
the contours are separated by a certain spacing
\cite{SchrammSheffield}.  But the techniques for analyzing the
double-dimer model \cite{kenyon:ddci} depend in an essential way upon
the lattice being bipartite, which the Fisher lattice is not.  Dimers
on the Fisher lattice have no height function, and the monomer
correlations for the Fisher lattice
\cite{PhysRevB.12.3908,PhysRevB.17.1464} behave quite differently from
monomer correlations on the square and hexagonal lattices
\cite{MR2172582}.  Nonetheless, as we shall see, individual
double-dimer loops on the critical Fisher lattice appear to have the
same limiting behavior as loops of the square- or hexagonal-lattice
double-dimer models, but there are $\sqrt{2}$ times fewer loops in the
Fisher-lattice double-dimer model.

The polarization of the double-Ising model (the XOR-Ising
configuration) at the critical temperature has been identified with a
``spin wave operator'' 
of the Gaussian free field \cite{kadanoff-brown} (see also
\cite{Knops1980448,hilhorst-leeuwen,dF-S-Z,boyanovsky,MR992372}), and this identification has been used to compute Ising
spin correlations \cite{PhysRevB.12.3908,PhysRevB.17.1464}.  However,
the focus was on correlation functions, and it is unclear how to use
these works to extract information about the geometric properties of
XOR-Ising interfaces.

Boutillier and de Tili\'ere \cite{boutillier-tiliere,tiliere} studied
the partition function of the double-Ising model on ``isoradial''
graphs, which include the hexagonal and square lattices.  They showed
that when the Ising models are critical, the partition function is
identical to the partition function for cycle-rooted spanning forests
on the same graph.  Spanning trees can be transformed into dimers on a
related bipartite graph \cite{temperley:tree,MR1756162}, which have a
height function whose scaling limit is the Gaussian free field.
However, there is no known algorithm for transforming cycle-rooted
spanning forests to pairs of Ising spin configurations, so it is not
clear how to use their results to relate the Gaussian free field to
XOR-Ising interfaces.

Sheffield and Werner gave a characterization of the conformal loop
ensemble $\CLE_\kappa$ for $8/3<\kappa\leq 4$.  The gasket
is the set of points not surrounded by any loop of $\CLE_\kappa$.
Sheffield and Werner showed that the $\CLE_\kappa$ gasket is the same
as the set of points not surrounded by any loop of the Brownian loop
soup with intensity $c=(3\kappa-8) (6-\kappa)/(2\kappa)$ (the central
charge) \cite{SheffieldWerner:Soup}.  Since $c(\kappa=3)=\frac12$ and
$c(\kappa=4)=1$, the intersection of two independent $\CLE_3$ gaskets
is a $\CLE_4$ gasket.  It follows immediately that the gasket for the
OR (rather than XOR) of two independent critical Ising models is the
same as the $\CLE_4$ gasket \cite{sheffield}, which is the scaling
limit of contour lines of the Gaussian free field.  It is not clear,
however, how to relate the OR-Ising loops to the XOR-Ising loops.

\section{XOR-Ising loops in the bulk}

We generated many XOR-Ising configurations on large $L\times L$ tori
and measured various properties of the XOR-Ising loop ensemble (see
\tbref{tbl:params}) whose values are known for $\CLE_\kappa$.  These
include the length of the longest loop (to measure the loop dimension
\cite{beffara}), the winding angle variance of the longest loop
\cite{WW,duplantier}, the average number of loops surrounding a point
\cite{CZ,SSW}, the variance in the number of loops surrounding a point
\cite{KW,SSW}, and the number of points surrounded by no loops (to
measure the gasket dimension \cite{D:vesicle,SSW}).  The first two
quantities are properties of individual loops, and match the values
for $\CLE_4$ loops.  The latter three quantities are properties of the
ensemble rather than individual loops, and these do \textit{not\/}
match the values for the $\CLE_4$ ensemble.  For example, the number
of XOR-Ising loops surrounding a point is smaller by a factor of
$\sqrt{2}$ than the number of $\CLE_4$ loops surrounding a disk of
radius $1/L$.

In view of the various connections between the double-Ising model and
the Gaussian free field, and the close match between the measured
properties of individual XOR-Ising loops and $\CLE_4$ loops, a natural
hypothesis is that the XOR-Ising loops are distributed according to
contour lines of the Gaussian free field, but with the height spacing
between contours larger than what it would be for $\CLE_4$ loops.  To
test this, we measured the electrical properties of the XOR-Ising
loops.

Consider the surface of a cylinder of length~$\ell\gg 1$, radius~$1$,
and girth~$2\pi$ made of a resistive medium with resistance $2\pi$.
Suppose there is a sequence of noncontractible loops
$\LL_1,\dots,\LL_k$ that wind around the short direction of the
cylinder.  Let $R_-(\LL_i)$ and $R_+(\LL_i)$ denote the electrical
resistances between loop $\LL_i$ and the left and right ends of the
cylinder respectively.  If the loops $\LL_1,\dots,\LL_k$ are $\CLE_4$
loops, then $R_-(\LL_{i+1})-R_-(\LL_{i})$ is distributed according to
the exit time of a standard Brownian motion from the interval
$[-\pi,\pi]$ \cite{KW,SSW}, and in \cite{KW} it was predicted that the
electrical thickness $\ell-(R_-(\LL_i)+R_+(\LL_i))$ of loops $\LL_i$
far from either end of cylinder is distributed according to the
\textit{excursion\/} time of a standard Brownian motion from the
interval $[-\pi,\pi]$.  Schramm and Sheffield proved that $\CLE_4$ is
the scaling limit of the contour lines of the Gaussian free field with
a certain spacing $\lambda$ between the contour heights
\cite{SchrammSheffield}.  It also follows from their work that when
the height spacing of the contours is $x \lambda$, then
$R_-(\LL_{i+1})-R_-(\LL_{i})$ is distributed according the exit time
of standard Brownian motion from the asymmetric interval $[-\pi,x\pi]$
\cite{SchrammSheffield,sheffield}.
For the measured XOR-Ising loop density in \tbref{tbl:params}, the
appropriate value of $x$ is $\sqrt{2}$; this value is also consistent
with the variance and gasket dimension measurements.

\begin{figure}[t]
\ \ \psfrag{thickness}[tl][cc]{\ \ \begin{tabular}{l}
Two cumulative distribution functions:\\
1) Electrical thickness of XOR-Ising loops\\
2) Excursion time of Brownian motion from $[-\pi,\pi]$
\end{tabular}}
\includegraphics[width=\columnwidth,height=0.5\columnwidth]{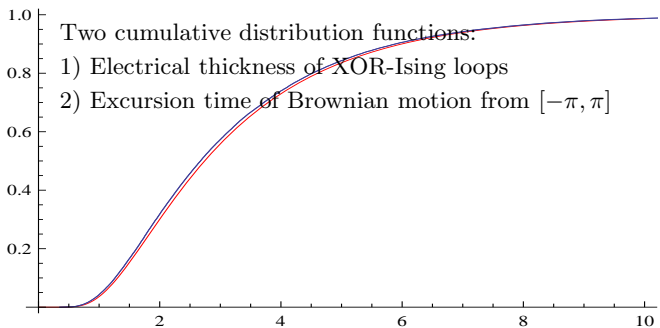}
\caption{The electrical thickness of XOR-Ising loops closely matches
  the Brownian excursion time from $[-\pi,\pi]$.}
\label{thickness}
\end{figure}

\begin{figure}[t]
\ \ \psfrag{radiidiff}[tl][cc]{\ \ \begin{tabular}{l}
Two cumulative distribution functions:\\
1) Difference in log-conformal radii of XOR-Ising loops\\
2) Exit time of Brownian motion from $[-\pi,\sqrt{2}\pi]$
\end{tabular}}
\includegraphics[width=\columnwidth,height=0.5\columnwidth]{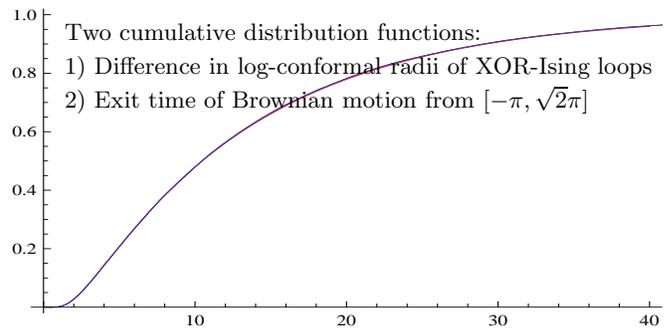}
\caption{The difference in log-conformal radii of XOR-Ising loops 
  closely matches the Brownian exit time from $[-\pi,\sqrt{2}\pi]$.}
\label{radii-diff}
\end{figure}

We generated many XOR-Ising loop configurations on $2^7\times 2^{15}$
cylinders, and using Marshall's Zipper program \cite{marshall-rohde}
computed the electrical resistances involving the noncontractible
loops.  As shown in Figures~\ref{thickness} and \ref{radii-diff},
there is excellent agreement between the measured distributions and
the distributions for contour lines of the GFF with height spacing
$\sqrt{2}$ times as large as for $\CLE_4$.

\section{XOR-Ising interfaces in\\ domains with boundary}

\begin{figure}[b]
\includegraphics[width=0.68\columnwidth]{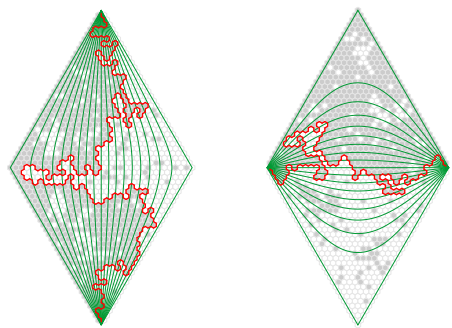}
\caption{(Color online) On the left is the XOR-Ising interface from
  bottom to top from \fref{fig:xor-ising}; on the right is the
  corresponding XOR-Ising interface from left to right.  When the
  domain is conformally mapped to the upper half plane with start- and end-points
  of the interface mapped to $0$ and $\infty$, the green curves get
  mapped to rays $r e^{i \theta}$ for constant $\theta$.  When
  plotting left-crossing probabilities in \fref{fig:left-plots}, for
  the region on the left we took sites on the short diagonal and
  compared the site's $\theta$ with the site's left-crossing probability;
  for the region on the right we did this for sites on the long
  diagonal.  }
\label{fig:long-short}
\end{figure}

\begin{figure}
\psfrag{solid}[tl][tc]{\ \begin{tabular}{l}
Three left-crossing probability curves:\\
1) XOR-Ising interface with\\[-2pt]
\phantom{1) }black-white / all-black b.c.\\
\phantom{1}a) connecting near opposite\\[-2pt]
\phantom{1a)} corners of rhombus\\
\phantom{1}b) connecting far opposite\\[-2pt]
\phantom{1b)} corners of rhombus\\
2) $\SLE_{4,\sqrt{2}-1,\sqrt{2}-1}$
\end{tabular}}
\psfrag{alternating}[tl][tc]{\ \begin{tabular}{l}
Three left-crossing probability curves:\\
1) XOR-Ising interface with\\[-2pt]
\phantom{1) }alternating / alternating (with phase shifts) b.c.\\
\phantom{1}a) connecting near opposite corners of rhombus\\
\phantom{1}b) connecting far opposite corners of rhombus\\
2) $\SLE_{4,-1,-1}$
\end{tabular}}
\psfrag{antiperiodic}[tl][tc]{\ \ \ \ \ \ \begin{tabular}{l}
Two left-crossing probability curves:\\
1) XOR-Ising interface from cylinder\\[-2pt]
\phantom{1) }with antiperiodic b.c.\\
2) $\SLE_{4,1/\sqrt{2}-1,1/\sqrt{2}-1}$
\end{tabular}}
\psfrag{+}[cc][cc][0.7]{$\oplus$}
\psfrag{=}[cc][cc][0.7]{$=$}
\includegraphics[width=\columnwidth]{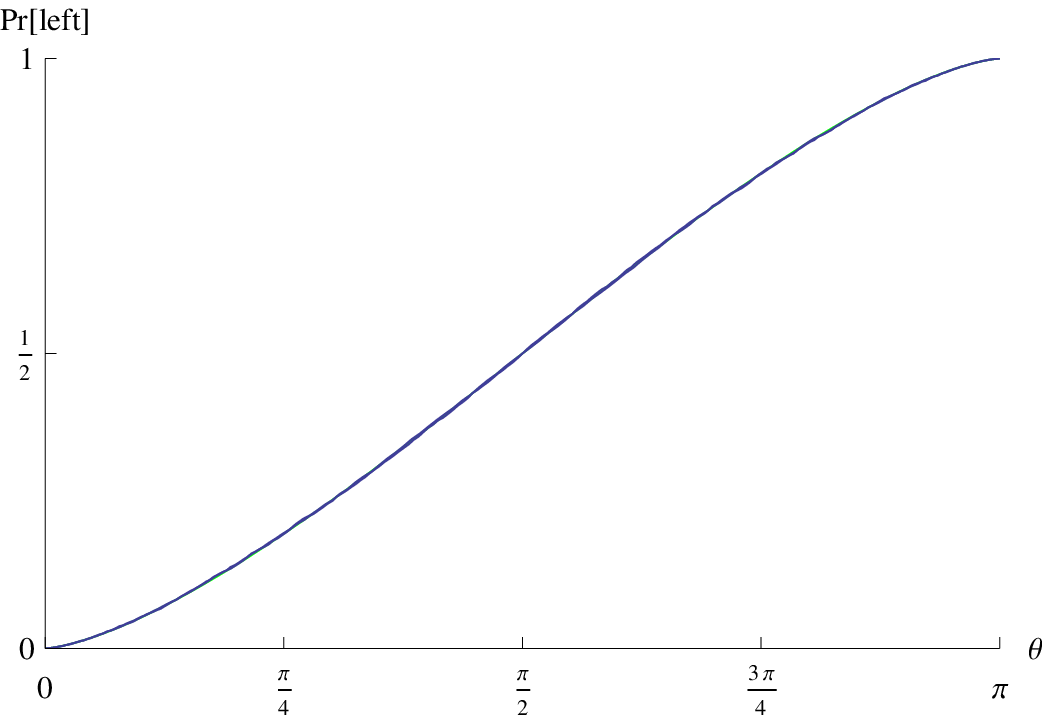}
\smash{\raisebox{30pt}{\hspace{100pt}\includegraphics[width=0.5\columnwidth]{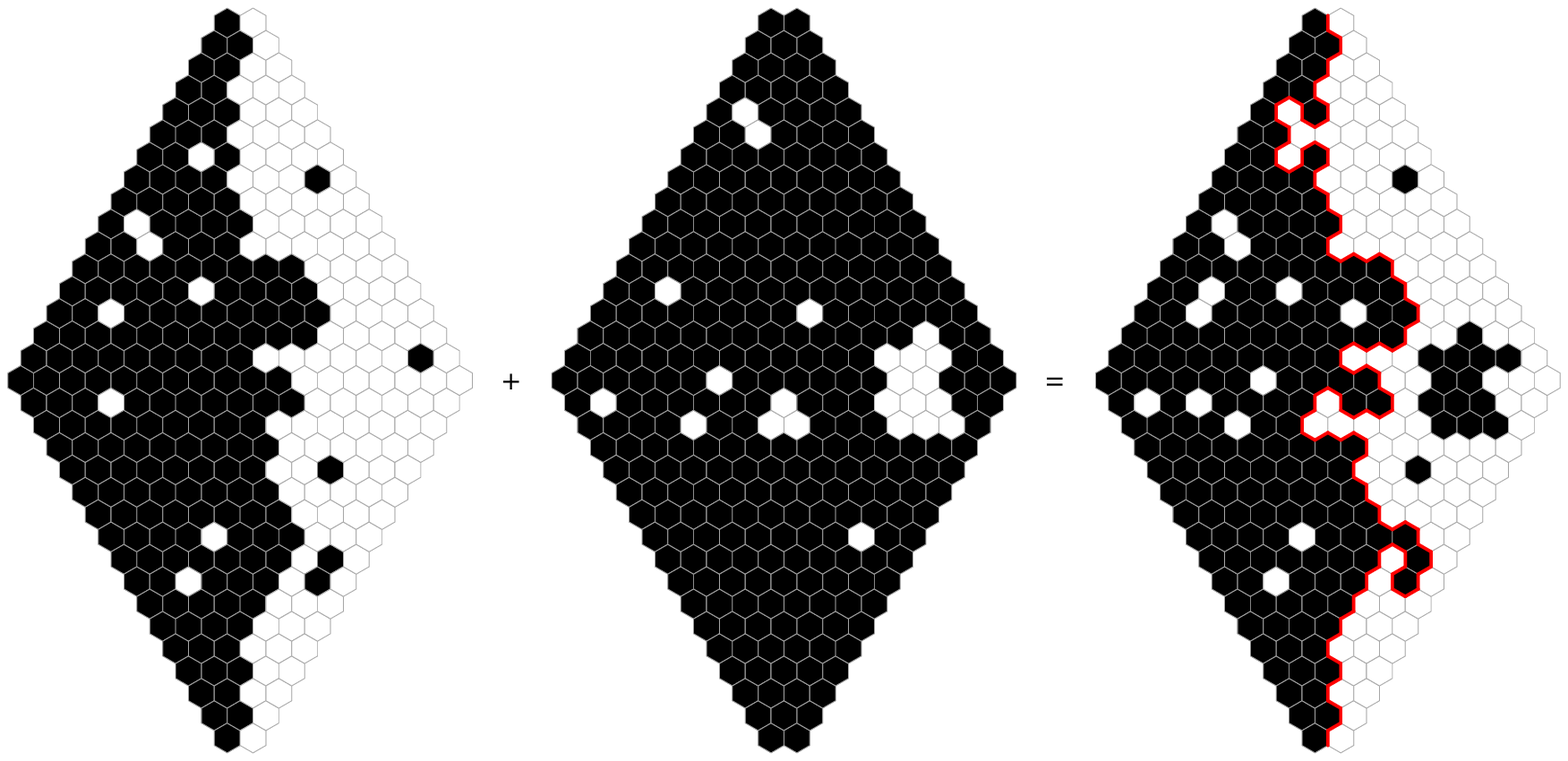}}}\vspace{-0.3pt}
\includegraphics[width=\columnwidth]{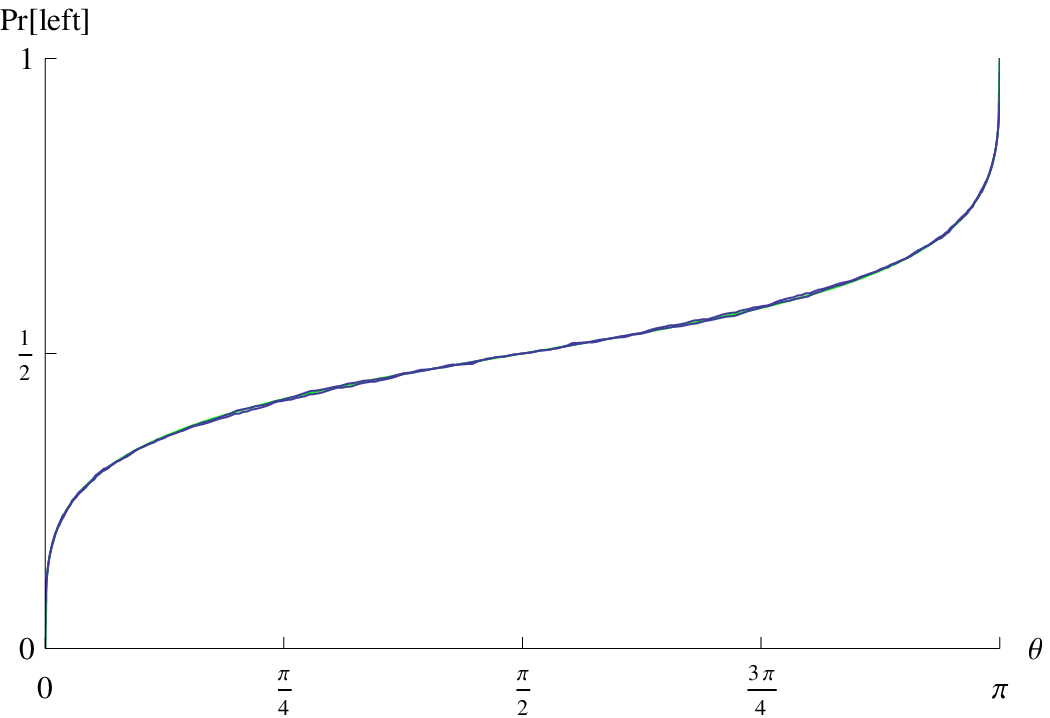}
\smash{\raisebox{30pt}{\hspace{100pt}\includegraphics[width=0.5\columnwidth]{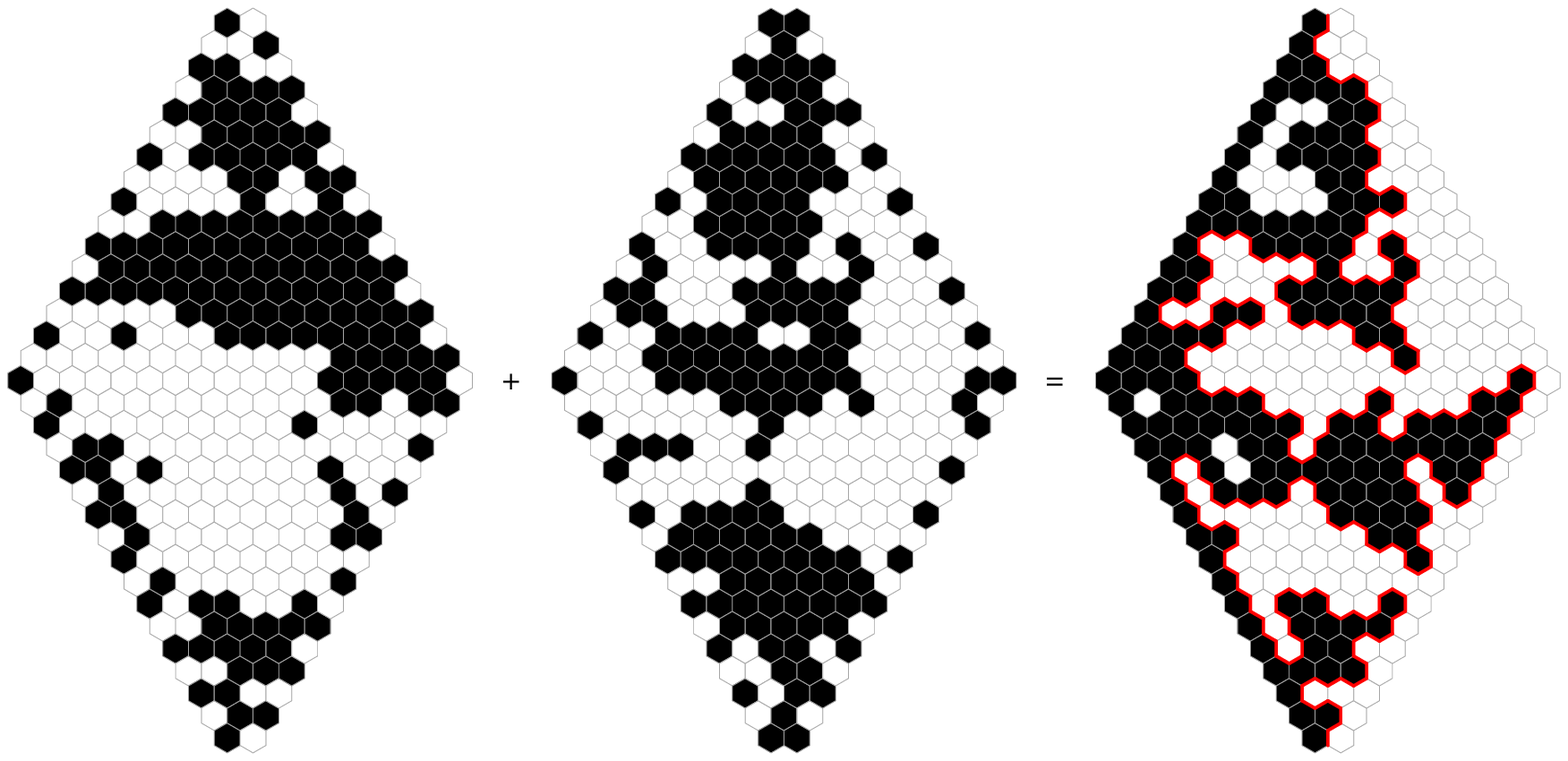}}}\vspace{-0.3pt}
\includegraphics[width=\columnwidth]{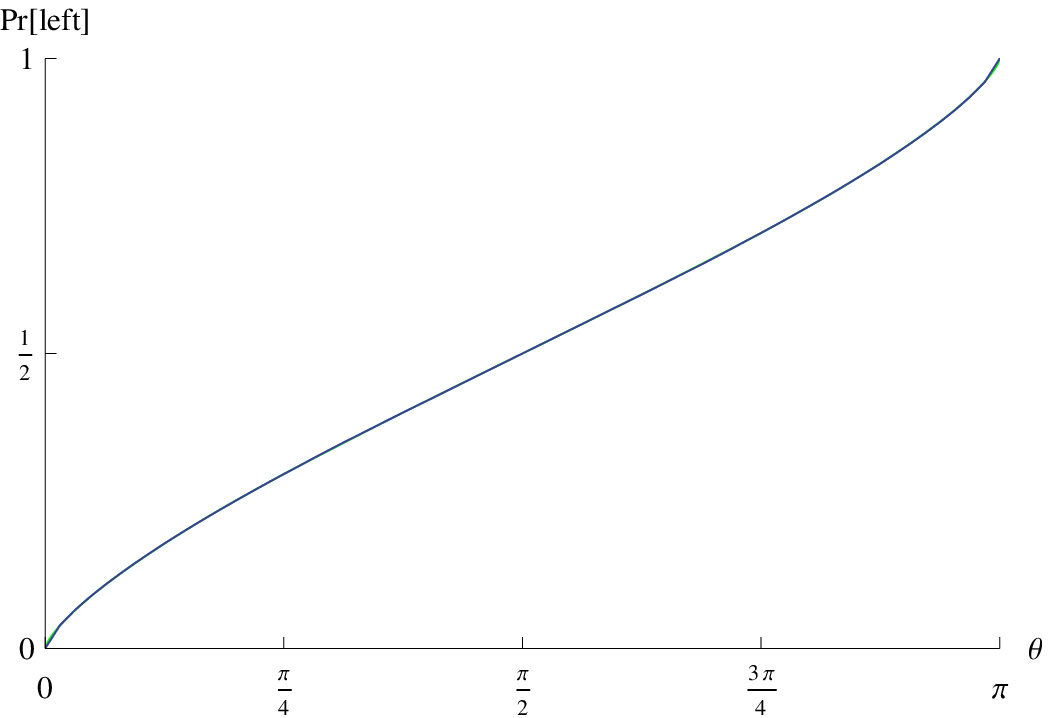}
\caption{(Color online) Left-crossing probabilities for XOR-Ising
  interfaces with different boundary conditions and different
  geometries, compared with the left-crossing probabilities for
  $\SLE_{4,\rho,\rho}$ for different $\rho$'s.  (The top, middle, and
  bottom plots show 3 curves, 3 curves, and 2 curves, respectively; in
  each plot the curves are essentially indistinguishable.)  The
  geometry of the domain appears not to matter, consistent with
  conformal invariance of the XOR-Ising interfaces.  Different
  boundary conditions correspond to different $\rho$'s.  The
  left-crossing probabilities for $\SLE_{4,\rho,\rho}$ were measured
  by sampling the left-crossing probabilities of a contour line in a
  Gaussian free field with $\pm (1+\rho) \lambda$ boundary conditions
  \cite{SchrammSheffield}.}
\label{fig:left-plots}
\end{figure}

To further test the connection between XOR-Ising interfaces and the
Gaussian free field, we measured left-crossing probabilities of the
XOR-Ising paths connecting two boundary points of a domain with
boundary.  \fref{fig:long-short} illustrates the coordinate $\theta$
against which we plotted left-crossing probabilities in
\fref{fig:left-plots}.  There are a variety of boundary conditions for
the two Ising models that force an interface in the XOR-Ising
configuration.  For the solid boundary conditions in
\fref{fig:xor-ising}, the left-crossing probabilities of the interface
are consistent with $\SLE_{4,\sqrt{2}-1,\sqrt{2}-1}$ (see
\fref{fig:left-plots}), which corresponds to boundary heights of
$\pm\sqrt{2}\lambda$ for the Gaussian free field \cite{SchrammSheffield}.
If instead the boundary spins alternate between black and white
(effectively free boundary conditions), with an interface forced by
two phase shifts (see Figure~\ref{fig:left-plots}), then the left-crossing
probabilities are consistent with $\SLE_{4,-1,-1}$, which corresponds
to boundary heights of $\pm\eps$ (in the limit $\eps\downarrow 0$) for the Gaussian free field.

We also tried a third set of boundary conditions, which specifies the
boundary spin values of neither Ising model, but instead conditions on
their forming an interface.  We did this by generating Ising spin
configurations on a long cylinder with antiperiodic boundary
conditions, and taking the two Ising spin configurations from the
upper half and lower half of the cylinder.  For XOR-Ising interfaces
obtained in this way, the left-crossing probabilities are consistent
with $\SLE_{4,1/\sqrt{2}-1,1/\sqrt{2}-1}$.  This is somewhat
surprising, since these boundary conditions satisfy the domain Markov
property at the discrete level, so one might expect the interface to
be simply $\SLE_4$.  But evidently these boundary conditions act
differently on smooth boundaries than they do on rough boundaries.

\begin{acknowledgments}
  We thank Scott Sheffield, Gady Kozma, and Bernard Nienhuis for
  useful discussions.
\end{acknowledgments}

\bibliography{xorising}

\end{document}